\documentstyle[prl,aps]{revtex}
\twocolumn

\begin{document}
\author{Jian Qi Shen $^{1,}$$^{2}$ \footnote{E-mail address: jqshen@coer.zju.edu.cn}}
\address{$^{1}$  Centre for Optical
and Electromagnetic Research, State Key Laboratory of Modern
Optical Instrumentation,\\ Zhejiang University,
Hangzhou Yuquan 310027, P.R. China\\
$^{2}$ Zhejiang Institute of Modern Physics and Department of
Physics, Zhejiang University, Hangzhou 310027, P.R. China}
\date{\today }
\title{A scheme of measurement of quantum-vacuum geometric phases \\in the noncoplanar fibre system}
\maketitle

\begin{abstract}
We study the {\it quantum-vacuum geometric phases} resulting from
the vacuum fluctuation of photon fields in Tomita-Chiao-Wu
noncoplanar curved fibre system, and suggest a scheme to test the
potential existence of such vacuum effect. Since the signs of the
quantum-vacuum geometric phases of left- and right- handed (LRH)
circularly polarized light are just opposite, the total geometric
phases at vacuum level is inescapably absent in the fibre
experiments performed previously by other authors. By using the
present approach where the fibre made of gyroelectric media is
employed, the quantum-vacuum geometric phases of LRH light cannot
be exactly cancelled, and may therefore be achieved test
experimentally.

\end{abstract}

Historically, since Berry's discovery of Berry (geometric) phase
in adiabatic quantum processes\cite{Berry}, geometric phase
problems had captured extensive attention of researchers in
various physical areas such as quantum optics\cite{Gong},
condensed matter physics\cite{Taguchi,Falci,Shenprb}, gravity
theory\cite {Furtado} as well as molecular physics (and molecular
reaction)\cite {Kuppermann}. Due to its topological and global
properties, recently, geometric phases was applied to some areas
such as quantum decoherence and geometric (topological) quantum
computation\cite{Wu,Jones,Wang,Zhusl}. More recently,
Fuentes-Guridi {\it et al}\cite{Fuentes} and we\cite{Shenpla}
investigated independently the geometric phases at quantum-vacuum
level. In the work of Fuentes-Guridi {\it et al}, they
investigated the geometric phases at vacuum level in the spin
system where a spin-1/2 particle interacts with a quantized
magnetic field, and they referred to such topological (geometric)
phases as the ``vacuum-induced geometric phases''\cite{Fuentes}.
In our work, we studied the similar vacuum geometric phases (which
was referred to as the ``quantum-vacuum geometric
phases'')\cite{Shenpla} in Tomita-Chiao-Wu noncoplanar fibre
system, where a second-quantized electromagnetic wave propagates
inside a noncoplanarly curved fibre\cite{Chiao,Tomita}. It was,
however, unfortunate that in the fibre system the {\it
quantum-vacuum geometric phases} of left- and right- handed (LRH)
circularly polarized light are just cancelled by each other and
that such a cancellation may lead this observable vacuum effect to
be absent experimentally. In order to investigate this physically
interesting geometric phases at quantum-vacuum level, here, we
will propose an experimentally feasible scheme by means of a
noncoplanarly curved fibre made of gyroelectric media. A
remarkable feature of the present experimental realization is that
by using such gyroelectric fibres, it is impossible for the
quantum-vacuum geometric phases of LRH light to be exactly
cancelled, which, therefore, will enable us to test and
experimentally investigate the quantum-vacuum geometric phases in
the noncoplanarly curved fibre system. Since quantum-vacuum
geometric phases has close relation to the vacuum fluctuation
energies, its experimental realizations may be relevant to a
fundamental topic, {\it i.e.}, the validity problem of
normal-order procedure for the operator products in time-dependent
quantum field theory, which will be considered in what follows and
discussed further in the conclusion of the present paper.

Even though many previous investigations treated the geometric
phases in the noncoplanarly curved fibre system by using the
differential geometrical method ({\it e.g.}, the concept of
parallel transport), Maxwellian theory and Berry's phase
formulation\cite{Kwiat,Robinson,Haldane1,Haldane2}, all these
studies may not be of second quantization and therefore might give
no insight into the geometric phases at quantum-vacuum level. In
order to consider the quantal character of geometric phases of
photons in the fibre system, one should investigate the
second-quantized spin model, which can characterize the motion of
photon moving inside the noncoplanarly curved
fibre\cite{Chiao,Tomita}. But if the normal-order procedure is
employed in the second-quantized spin model ({\it i.e.}, the
photon spin operator is of normal order), then the quantum-vacuum
geometric phases will inevitably be removed. This may be
understood as follows: the quantum-vacuum geometric phases of
photons results from the zero-point energy of vacuum quantum
fluctuation, which will be removed by the normal-order procedure,
just as the case in time-independent quantum field
theory\cite{Bjorken}. Such a technique in {\it time-independent}
quantum field theory is harmless since we remove the same amount
of background energy at different time ({\it i.e.}, the vacuum
background energy is removed {\it globally}). However, in the {\it
time-dependent} field theory, if we use the normal-order
procedure, we will remove the different amount of background
energy at different time in the evolution process ({\it i.e.}, the
vacuum background energy is removed based on the different
cardinal numbers at different time). This, therefore, means that
the normal-order procedure will unavoidably remove some potential
physical effects (such as the quantum-vacuum geometric phases) of
vacuum of time-dependent quantum systems, since the physically
interesting geometric phases arises in the time-dependent systems
(or the systems whose Hamiltonians possess the evolution
parameters). So, to consider the quantum-vacuum geometric phases
of photons inside the coiled optical fibre, we should deal with
the second-quantized spin model, the Hamiltonian of which is of
non-normal order. By using the Lewis-Risenfeld invariant
theory\cite{Riesenfeld}, one can obtain the photon wavefunction
inside the curved fibre as follows\cite{Shenpla}
\begin{equation}
|\Psi(t)\rangle=\exp\left[\frac{1}{i\hbar}\phi(t)\right]V(t)|n_{\rm
R}, n_{\rm L}\rangle       \label{eq1}
\end{equation}
with $V(t)=\exp [\beta (t)S_{+}-\beta ^{\ast }(t)S_{-}]$, where
$S_{+}$ and $S_{-}$ are the photon spin operators, and the
time-dependent parameters $\beta (t)=-\frac{\theta (t)}{2}\exp
[-i\varphi (t)]$, $\beta ^{\ast }(t)=-\frac{\theta(t)}{2}\exp
[i\varphi(t)]$. Here the polar angle $\theta$ and the azimuthal
angle $\varphi$ are so defined that ${\bf k}=k(\sin \theta \cos
\varphi, \sin \theta \sin \varphi, \cos \theta)$ denotes the wave
vector of photon inside the curved fibre. The noncyclic
nonadiabatic geometric phase $\phi(t)$ in (\ref{eq1}) is of the
form
\begin{equation}
\phi(t)=\left[\left(n_{\rm R}+\frac{1}{2}\right)-\left(n_{\rm
L}+\frac{1}{2}\right)\right]\phi_{0}(t),       \label{eq2}
\end{equation}
where
$\phi_{0}(t)=\int_{0}^{t}\dot{\varphi}(t')\left[1-\cos\theta(t')\right]{\rm
d}t'$ (with dot denoting the derivative with respect to time $t$),
and $n_{\rm R}$ and $n_{\rm L}$ denote the occupation number
operators of right- and left- handed circularly polarized light,
respectively. It follows from Eq.(\ref{eq2}) that
\begin{equation}
\phi(t)=\left(n_{\rm R}-n_{\rm L}\right)\phi_{0}(t). \label{eq3}
\end{equation}
Although it seems that the expression (\ref{eq2}) is
mathematically equivalent to (\ref{eq3}), there is an essential
difference in physical meanings between them, namely, the
expression (\ref{eq2}) contains the quantum-vacuum geometric
phases, while the expression (\ref{eq3}) has no such vacuum
effect. In other words, the expression (\ref{eq3}) is in
connection only with geometric phases at {\it quantum} level
rather than at {\it quantum-vacuum} level. In the
papers\cite{Zhu,Gao2}, the authors dealt with the expression
(\ref{eq3}) for the photon geometric phases derived via the
normal-order procedure, which has removed the vacuum effect even
at the beginning of their calculations. Note, however, that here
the expression (\ref{eq3}) is derived from (\ref{eq2}), {\it
i.e.}, the non-normal-order Hamiltonian describing the motion of
photons in the coiled fibre is involved in the calculation of
geometric phases (\ref{eq2}). So, the absence of the
quantum-vacuum geometric phases $ \phi_{\rm R}^{\rm
vac}(t)=+\frac{1}{2}\phi_{0}(t)$, $ \phi_{\rm L}^{\rm
vac}(t)=-\frac{1}{2}\phi_{0}(t) $ in (\ref{eq3}) is due to the
fact that the signs of $\phi_{\rm R}^{\rm vac}(t)$ and $\phi_{\rm
L}^{\rm vac}(t)$ are just opposite (not due to the normal-order
procedure). Apparently, this cancellation makes the concept of
{\it quantum-vacuum geometric phases} trivial in physical
meanings, and such vacuum effect would not be observed in the
previous fibre experiments\cite{Tomita,Robinson}.

In what follows we will propose a new scheme which can detect the
above vacuum effect in the fibre system. Let us consider the wave
propagation in the gyroelectric media, the electric permittivity
tensor of which is as follows\cite{Veselago}
\begin{equation}
\hat{\epsilon}=\left({\begin{array}{cccc}
{\epsilon_{1}}  & {i\epsilon_{2} }& {0} \\
{-i\epsilon_{2}} &   {\epsilon_{1}} & {0}  \\
 {0} &  {0} &  {\epsilon_{3}}          \\
 \end{array}   }
 \right).                                  \label{eq5}
\end{equation}
If the wave vector of the electromagnetic wave propagating inside
the gyroelectric medium (with the magnetic permeability $\mu=1$)
is parallel to the third component of 3-D Cartesian coordinate
system, then the optical refractive indices squared takes the form
$n_{\pm}^{2}=\epsilon_{1}\pm \epsilon_{2} $\cite{Veselago,arxiv},
where $n_{+}$ and $n_{-}$ correspond to the right- and left-
handed circularly polarized light, respectively. In the following
we will analyze the motion of photons inside the coiled fibre made
of the above gyroelectric medium. For simplicity, here we only
consider the case of photon precessional frequency on the fibre
helicoid $\dot{\varphi}=\Omega$ (which is constant) and nutational
frequency $\dot{\theta}=0$ ({\it i.e.}, $\theta$ is constant).
Thus the noncyclic quantum-vacuum geometric phases of right- and
left- handed polarized light are of the form
\begin{equation}
\phi^{\rm vac}_{\rm R}(t)=+\frac{1}{2}\Omega_{+}(1-\cos\theta)t,
\quad      \phi^{\rm vac}_{\rm
L}(t)=-\frac{1}{2}\Omega_{-}(1-\cos\theta)t,
\end{equation}
respectively, where the corresponding precessional frequencies of
photon moving on the helicoid are given
\begin{equation}
\Omega_{+}=\frac{2\pi c}{\sqrt{d^{2}+(4\pi
a)^{2}}}\frac{1}{n_{+}},   \quad     \Omega_{-}=\frac{2\pi
c}{\sqrt{d^{2}+(4\pi a)^{2}}}\frac{1}{n_{-}}.    \label{eq8}
\end{equation}
Here $d$ and $a$ respectively denote the pitch length and the
radius of the helix of the noncoplanar fibre, and $c$ is the speed
of light in a vacuum. Hence, the total noncyclic quantum-vacuum
geometric phases of LRH light in the curved fibre system is
\begin{eqnarray}
\phi^{\rm vac}_{\rm tot}(t)&=&\phi^{\rm vac}_{\rm R}(t)+\phi^{\rm
vac}_{\rm L}(t)         \nonumber \\
&=&\frac{n_{-}-n_{+}}{n_{+}n_{-}}\frac{\pi c}{\sqrt{d^{2}+(4\pi
a)^{2}}}(1-\cos\theta)t,
\end{eqnarray}
which is no longer vanishing, and may be tested experimentally.

With the appropriate choice of the parameters $\epsilon_{1}$,
$\epsilon_{2}$ in Eq.(\ref{eq5}) which leads to the condition,
say, $n_{-}\gg n_{+}$, it follows from Eq.(\ref{eq8}) that the
vacuum geometric phase of left-handed polarized light may be
ignored and the only retained cyclic vacuum geometric phase is
\begin{equation}
\phi^{\rm vac}_{\rm tot}(T)\simeq\phi^{\rm vac}_{\rm
R}(T)=\frac{1}{2}\cdot 2\pi(1-\cos\theta)       \label{eq9}
\end{equation}
with $T=\frac{2\pi}{\Omega_{+}}=\frac{n_{+}\sqrt{d^{2}+(4\pi
a)^{2}}}{c}$. In the expression (\ref{eq9}), the coefficient
factor $\frac{1}{2}$ of the solid angle $2\pi(1-\cos\theta)$
subtended at the centre by a curve traced by the photon wave
vector results from the vacuum nature and characterizes the
quantum-vacuum feature of cyclic geometric phase in (\ref{eq9}).

It is clear that in order to detect the quantum-vacuum geometric
phases, in the fibre experiment one should have a gyroelectric
fibre with suitable electromagnetic parameters $\epsilon_{1}$,
$\epsilon_{2}$ of electric permittivity tensor. This may be a
challenging problem until now. During the past 20 years, the
design and fabrication of artificial materials such as artificial
chiral materials, photonic crystals and left-handed media
attracted intensive attention in various scientific and
technological areas\cite{Ziolkowski}. More recently, the fibres
fabricated from the artificial (anisotropic) composite materials
such as the left-handedness fibres and photonic crystal fibres are
studied both theoretically and experimentally\cite{Knight}. It may
be believed that if the investigation of the above
artificial-material fibres achieves success in experiments, it
might also be possible to design and fabricate the gyroelectric
fibre made of the anisotropic gyroelectric materials.

The quantum-vacuum geometric phases found here possesses
interesting properties, for it has an important connection with
the topological nature of time evolution of quantum vacuum
fluctuation. Fuentes-Guridi {\it et al.} and we think that in a
strict sense, the Berry phase has been studied only in a
semiclassical context\cite{Fuentes}, and that less attention was
paid to such geometric phases at purely quantum-vacuum level.
According to Fuentes-Guridi {\it et al.}'s statement, such vacuum
geometric phases may open up a new areas to the study of the
consequences of field quantization in the geometric evolution of
states\cite{Fuentes}, so, it is emphasized that the vacuum effect
presented here deserves further consideration both theoretically
and experimentally.

In this paper we propose an experimental scheme to test the
geometric phases at quantum-vacuum level by using the noncoplanar
gyroelectric fibre. Here the expression (\ref{eq2}) for the
quantum-vacuum geometric phases is derived without the
normal-order procedure. In the conventional time-independent field
theory, the normal-order procedure is useful to remove the
divergence of vacuum energy and charge. However, to the best of
our knowledge, in the literature, the validity problem of
normal-order procedure in time-dependent quantum field theory gets
less attention than it deserves. The physical interesting
geometric phases arises in the time-dependent quantum system.
However, its vacuum contribution may be removed by the
normal-order technique. So, we think that it is necessary to
consider the problem as to whether the normal-order procedure is
valid or not in time-dependent quantum field theory. According to
the above discussion, detection of the existence of vacuum
geometric phases may deal with this problem in time-dependent
quantum field theory. If experimental evidences reveal that there
truly exists the vacuum geometric phases, then it will be
reasonably believed that the normal-order technique might be no
longer valid in time-dependent quantum field theory. But if we
fail to detect the vacuum geometric phases experimentally ({\it
i.e.}, there exists no such vacuum geometric phases), then we may
argue that the normal-order technique is still valid in
time-dependent quantum field theory. We hope the quantum-vacuum
geometric phases would be investigated experimentally by using the
present scheme in the near future.

\textbf{Acknowledgements}  This project was supported by the
National Natural Science Foundation of China under Project No.
$90101024$ and $60378037$.

\end{document}